\documentclass[conference]{IEEEtran}
\IEEEoverridecommandlockouts
\usepackage{cite}
\usepackage{amsmath,amssymb,amsfonts}
\usepackage{algorithm}
\usepackage[printonlyused,nolist]{acronym}
\usepackage{algpseudocode}
\usepackage{graphicx}
\usepackage{textcomp}
\usepackage{xcolor}
\usepackage{adjustbox}
\usepackage[version=4]{mhchem}
\usepackage{booktabs}
\usepackage{multirow}
\usepackage{siunitx}
\usepackage{eurosym}
\usepackage{mhchem}
\usepackage{amssymb}
\usepackage{tikz}
\usepackage{pgfplots}
\pgfplotsset{compat=1.7}
\usepackage{pgfplotstable}
\usepgfplotslibrary{fillbetween}
\pgfdeclarelayer{bg}    % Background layer
\pgfdeclarelayer{main}  % Main layer
\usepgfplotslibrary{groupplots} % Load the groupplots library
\usepackage{multirow}
\usepackage{graphicx}
\usepackage{hyperref}
\usepackage{eurosym}
\usepackage{subcaption}
\def\BibTeX{{\rm B\kern-.05em{\sc i\kern-.025em b}\kern-.08em
    T\kern-.1667em\lower.7ex\hbox{E}\kern-.125emX}}
\usepackage{url}

\newtheorem{definition}{Definition} % Define a new theorem environment for definitions

\begin{document}

%\title{Distributionally Robust Chance-Constrained Optimization for  Flexible Resources Bidding in Nordic Ancillary Service Markets}

\title{Leveraging $\rm{P90}$ Requirement: Flexible Resources Bidding in Nordic Ancillary Service Markets}

\author{Peter A.V. Gade\textsuperscript{*}\textsuperscript{\textdagger}, Henrik W. Bindner\textsuperscript{*}, Jalal Kazempour\textsuperscript{*} \\
    \textsuperscript{*}Department of Wind and Energy Systems, Technical University of Denmark, Kgs. Lyngby, Denmark \\
    \textsuperscript{\textdagger}IBM Client Innovation Center, Copenhagen, Denmark
    % <-this % stops a space
    \thanks{
        %Corresponding author. Tel.: +45 24263865. \\
        Email addresses: pega@dtu.dk (P.A.V. Gade), hwbi@dtu.dk (H.W. Bindner), and jalal@dtu.dk (J. Kazempour). The authors would like to thank Trygve Skjøtskift (IBM Client Innovation Center) and Thomas Dalgas Fechtenburg (Energinet) for discussions. The authors would also like to acknowledge the financial support from Innovation Fund Denmark under grant number 0153-00205B.}% <-this % stops a space
    \vspace{-3mm}
}

% \footnote{Corresponding author. Tel.: +45 24263865. \\ Email addresses: pega@dtu.dk (P.A.V. Gade), jalal@dtu.dk (J. Kazempour).}

%\markboth{Journal of \LaTeX\ Class Files,~Vol.~14, No.~8, August~2021}%{Shell \MakeLowercase{\textit{et al.}}: A Sample Article Using IEEEtran.cls for IEEE Journals}


\maketitle

% \tableofcontents

\IEEEaftertitletext{\vspace{-0.8\baselineskip}}
\maketitle
\thispagestyle{plain}
\pagestyle{plain}
\begin{abstract}
The $\rm{P90}$ requirement of the Danish transmission system operator, Energinet, incentivizes flexible resources with stochastic power consumption/production baseline to bid in Nordic ancillary service markets with the minimum reliability of 90\%, i.e., letting them cause reserve shortfall with the probability of up to 10\%. Leveraging this requirement, we develop a distributionally robust joint chance-constrained optimization model for aggregators of flexible resources to optimize their volume of reserve capacity to be offered. Having an aggregator of electric vehicles as a case study, we show how distributional robustness is key for the aggregator when making bidding decisions in a non-stationary uncertain environment. We also develop a heuristic based on a grid search for the system operator to adjust the $\rm{P90}$ requirement and the level of conservativeness, aiming to procure the maximum reserve capacity from stochastic resources with least expected shortfall.  
%Stochastic flexible resources can now offer their capacity to Nordic ancillary services under a new regulation that addresses their stochastic nature. In this work, we show how aggregators can exploit the market regulations by generalizing the requirement in a mathematical representation that is distributionally robust with respect to their uncertain offering capacity. A tractable formulation is provided in the form of a MILP, and it is agnostic to the underlying technology offering its flexibility. We show how a TSO can maximize its available flexible capacity using a bi-level optimization problem where the inner problem represents the aggregators' profit maximization. This is illustrated in a simulated case study of an aggregator with a portfolio of electric vehicles. 
\end{abstract}

\vspace{2mm}


% ############## OBS: MAX 10 PAGES!!!! ##############

\vspace{-2mm}

% TODO: delete
% \section{Acronyms}\label{sec:acronyms}

\begin{acronym}[DRJCCP]
    \acro{IoT}{Internet of Things}
    \acro{OPF}{Optimal Power Flow}
    \acro{IEEE}{Institute of Electrical and Electronic Engineers}
    \acro{CHP}{Combined Heat \& Power}
    \acroplural{CHP}[CHPs]{Combined Heat \& Power}
    \acro{SOCP}{Second-Order Cone Program}
    \acroplural{SOCP}[SOCPs]{Second-Order Cone Programs}
    \acro{MILP}{Mixed-Integer Linear Program}
    \acro{MINLP}{Mixed-Integer Nonlinear Program}
    \acro{LP}{Linear Program}
    \acro{AC}{alternating current}
    \acroplural{LP}[LPs]{Linear Programs}
    \acro{COP}{Coefficient of Performance}
    \acro{KKT}{Karush-Kuhn-Tucker}
    \acro{DSO}{Distribution System Operator}
    \acroplural{DSO}[DSOs]{Distribution System Operator}
    \acro{TSO}{Transmission System Operator}
    \acroplural{TSO}[TSOs]{Transmission System Operators}
    \acro{DER}{Distributed Energy Resource}
    \acroplural{DER}[DERs]{Distributed Energy Resources}
    \acro{TCL}{Thermostatically Controlled Load}
    \acroplural{TCL}[TCLs]{Thermostatically Controlled Loads}
    \acro{BRP}{Balance Responsible Party}
    \acroplural{BRP}[BRPs]{Balance Responsible Parties}
    \acro{NEMO}{Nominated Electricity Market Operator}
    \acroplural{NEMO}[NEMOs]{Nominated Electricity Market Operators}
    \acro{FCR}{Frequency Containment Reserve}
    \acro{FCR-N}{Frequency Containment Reserve - Normal Operation}
    \acro{FCR-D}{Frequency Containment Reserve - Disturbance}
    \acro{FFR}{Fast Frequency Reserve}
    \acro{mFRR}{Manual Frequency Restoration Reserve}
    \acro{aFRR}{Automatic Frequency Restoration Reserve}
    \acro{ENTSO-E}{European Network of Transmission System Operators for Electricity}
    \acro{TERRE}{Trans-European Replacement Reserves Exchange}
    \acro{MARI}{Manually Activated Reserves Initiative}
    \acro{PICASSO}{Platform for the International Coordination of Automated Frequency Restoration and Stable System Operation}
    \acro{ODE}{Ordinary Differential Equation}
    \acroplural{ODE}[ODEs]{Ordinary Differential Equations}
    \acro{PDE}{Partial Differential Equation}
    \acroplural{PDE}[PDEs]{Partial Differential Equations}
    \acro{SDE}{Stochastic Differential Equation}
    \acroplural{SDE}[SDEs]{Stochastic Differential Equations}
    \acro{NODE}{Neural Ordinary Differential Equation}
    \acroplural{NODE}[NODEs]{Neural Ordinary Differential Equations}
    \acro{PINN}{Physics-Informed Neural Network}
    \acroplural{PINN}[PINNs]{Physics-Informed Neural Networks}
    \acro{PID}{Propportional Integral Derivative}
    \acro{MPC}{Model Predictive Control}
    \acro{E-MPC}{Economic Model Predictive Control}
    \acro{OD}{Opening Degree}
    \acro{IS}{in-sample}
    \acro{OOS}{out-of-sample}
    \acro{CC}{Chance Constraint}
    \acroplural{CC}[CCs]{Chance Constraints}
    \acro{JCC}{Joint Chance Constraint}
    \acroplural{JCC}[JCCs]{Joint Chance Constraints}
    \acro{DRJCC}{Distributionally Robust Joint Chance Constraint}
    \acroplural{DRJCC}[DRJCCs]{Distributionally Robust Joint Chance Constraints}
    \acro{DRJCCP}{Distributionally Robust Joint Chance-Constrained Program}
    \acro{EV}{Electric Vehicle}
    \acroplural{EV}[EVs]{Electric Vehicles}
    \acro{CVaR}{Conditional Value at Risk}
    \acro{NP}{Non-deterministic Polynomial-time}
    \acro{SAA}{Sample Average Approximation}
    \acro{EV}{Electric Vehicle}
    \acroplural{EV}[EVs]{Electric Vehicles}
    \acro{LER}{Limited Energy Reservoir}
    \acroplural{LER}[LERs]{Limited Energy Reservoirs}
    \acro{PV}{Photovoltaic}
    \acroplural{PV}[PVs]{Photovoltaics}
\end{acronym}

\vspace{2mm}
\section{Introduction}\label{sec:Introduction}
\vspace{-1mm}

% \ac{DRJCC} for CCH EVs Nordic ancillary service markets. Focus is not on P90 or EV/technology

Flexible resources with stochastic electric power production or consumption have the capability to balance the power grid. Examples of such assets are  wind turbines in the supply side and electric vehicles and heat pumps in the demand side. However, their stochastic nature is a barrier to become ancillary service providers like other conventional units. The current requirements of Nordic ancillary service markets  technically allow stochastic resources to bid their flexibility. Leveraging such requirements, we develop a bidding decision-making problem for a portfolio of flexible resources, mathematically represented as a  \ac{DRJCCP}, optimizing the volume of reserve capacity to be offered. From the  perspective of flexibility aggregators, the proposed methodology is general, meaning it can be exploited for bidding decision making irrespective of the type and technology of  assets in their portfolio. %We show how aggregators with stochastic assets can optimally bid in Nordic ancillary service markets. 
In addition, we take the perspective of a \ac{TSO} and present a trade-off between increased  supply of ancillary services from stochastic flexible resources and uncertainty of delivery. 
%This is of a particular interest for the \ac{TSO}, while from an aggregator perspective, the methodology generalizes bidding as being either distributionally robust or empirically robust.

%\footnote{For the remainder of this paper, we refer to an aggregator as any entity that offers flexible capacity from at least one technology or asset into some ancillary service. Aggregator is thus used interchangeably with flexible provider.}

\vspace{2mm}
\subsection{Background: Nordic ancillary service market requirements}\label{sec:background}
\vspace{-1mm}
There has been an intention to make Nordic ancillary service markets attractive for stochastic resources to offer their flexibility. This is mainly due to two reasons: First, the Nordic power grid is undergoing a rapid transition towards least fossil-fuel based generation and more intermittent generation from wind and solar power, thus stressing the power grid. Second, the need of Nordic \acp{TSO} for frequency-related ancillary services has been increasing with a particular focus on flexibility procurement from green technologies, e.g., batteries, renewables, and flexible demands \cite{energinet2}. 

The Danish \ac{TSO}, Energinet, has recently published an innovative list of requirements \cite{energinet} that specifically addresses how stochastic  resources are pre-qualified for offering their flexibility into Nordic ancillary service markets. Among others, one requirement takes into account the stochastic nature of  flexible resources with respect to their unknown future flexible capacity. Currently, bidding to Nordic ancillary service markets generally occurs in the day-ahead stage when the \ac{TSO} books certain amounts of various ancillary services. Stochastic resources do not necessarily know with certainty their baseline power consumption/production 12-36 hours in advance. Hence, Energinet allows for a \textit{probability} of 10\% that the submitted bid in the day-ahead stage is not fully available in real time. 
This means that Energinet allows for reserve shortfall (also called overbidding) with the maximum probability of 10\%. Hereafter, we refer to it as ``the $\rm{P90}$ requirement" as it aims for bids with the reliability of at least 90\%. Energinet may check ex-post the realized frequency of reserve shortfall during a certain time period, e.g., three months. In case it goes beyond the allowed threshold, the corresponding flexibility provider may lose its qualification to bid in the market, and therefore should go through again the pre-qualification process of Energinet. A detailed definition of the $\rm{P90}$ requirement is provided later in the following section. 

This requirement naturally leads to a mathematical formulation for optimally offering flexible capacity using a chance-constrained program. An aggregator can exploit this requirement to its advantage by maximizing profits within the allowed violation range. However, this might not be in the aggregator's or the \ac{TSO}'s best interest as shown later. Any mis-specification of the aggregator's empirical distribution for offering  capacity might cause a violation of the $\rm{P90}$ requirement and result in less available flexible capacity for the \ac{TSO} in reality. We therefore investigate how a distributionally robust representation of the $\rm{P90}$ requirement can benefit aggregators when bidding capacity, while the \ac{TSO} can use this information to assess how well the $\rm{P90}$ requirement functions. For distributional robustness, we use the Wasserstein distance as a metric to represent a given \textit{conservativeness} when offering flexible capacity.

\vspace{2mm}
\subsection{Research questions and our contributions}
\vspace{-1mm}
%\acp{TSO} are looking to integrate more flexibility from non-fossil fuel based technologies, and flexible demand or production have the potential to provide some of the supply to meet the \ac{TSO} demand. 
%n order to integrate stochastic flexible resources, aggregators and \acp{TSO} face a number of challenges, respectively: 
From the perspective of an aggregator, the main question is how to develop a DRJCCP complying to the $\rm{P90}$ requirement and what level of conservativeness (e.g., in terms of Wasserstein distance) to be chosen when bidding into the Nordic ancillary service markets. From the the \ac{TSO}'s perspective, a relevant question is 
how the $\rm{P90}$ requirement and the conservativeness of aggregators impact the total reserve capacity supply from stochastic resources as opposed to the increased uncertainty of delivery. % and \textit{(iii)} increased supply, and therefore liquidity, decreases \ac{TSO} procurement costs.

To answer the first question from an aggregator's perspective, we formulate a bidding problem as a \ac{DRJCCP}, complying to the $\rm{P90}$ requirement, with the objective of profit maximization.  
As an example of a stochastic flexible resource, we consider a portfolio of \acp{EV}, whose consumption  is stochastic. We do not consider the option of EVs injecting to the grid, but rather adjusting their consumption level up and down when charging. 
Note that the proposed formulation can be applied to any other types of stochastic flexible resources. 
%The monetary value of offering Nordic ancillary services at different conservativeness levels is demonstrated by adjusting the Wasserstein distance. In particular, 
We show how the conservativeness level directly impacts the aggregator's security of supply in a non-stationary environment.

To answer the second question from a TSO's perspective, we show how a trade-off exists between increased supply of flexibility  and uncertainty of delivery. The find an optimal trade-off, we develop a bi-level optimization problem where the outer problem represents the \ac{TSO} setting the requirements in terms of allowed rate of shortfall reserve and conservativeness level. The inner problem represents an aggregator maximizing its profit using the proposed \ac{DRJCCP}. Thus, we assume the \ac{TSO} is able to not only adjust the $\rm{P90}$ requirement, but also demand stochastic flexible resources to consider a  certain level of conservativeness for robustness when offering their flexibility.

%We do not consider any specific market (or prices) and thus leave the third research question for future work. We do assume, however, that flexible providers are operationally and technically capable of responding, e.g., that a portfolio of \acp{EV} can deliver a frequency response. Furthermore, activation of any flexibility is assumed to contain negligible energy delivery, thus simplifying the objective of aggregators and \acp{TSO} by ignoring balancing energy. Hence, our work is primarily applicable to Nordic frequency markets, but agnostic to any stochastic flexible resource.

\vspace{2mm}
\subsection{Status quo and paper organization}
\vspace{-1mm}
A plethora of studies in the literature have investigated how stochastic flexible resources can participate in various ancillary service markets, both with respect to flexible demand  \cite{bondy2016procedure, bondy2014performance, biegel2014integration, AchievingControllabilityofElectricLoads} and flexible production from intermittent resources \cite{hansen2016provision, ullah2009wind, morey2023comprehensive, alshehri2019modelling}. However, no studies have looked specifically into the Nordic ancillary service market regulations, i.e., the $\rm{P90}$ requirement, when optimizing the bidding decisions of stochastic flexible resources. Reference \cite{zhang2018data} investigates how distributed energy resources can offer their flexible capacity using a chance-constrained program, but does not consider any market regulation like the $\rm{P90}$ requirement. There are also several papers that use distributionally robust optimization for bidding decisions. A distributionally robust bidding strategy for a wind-storage aggregator is proposed in \cite{Hug}. In addition, \cite{pierre} develops a distributionally robust optimization for wind energy trading under a two-price imbalance scheme. To the best of our knowledge, this is the first paper in the literature that develops a \ac{DRJCCP} for optimizing bidding decisions of stochastic flexible resources in ancillary service markets taking into account real market regulations.

%Other studies have extensively used (joint) chance-constrained programs from a TSO's or a market operator's perspective. Reference \cite{guo2020chance} shows how a joint energy and reserve market can be cleared, formulated as a chance-constrained program. A similar approach is taken in \cite{bienstock2014chance} for an optimal power flow problem. In \cite{roald2016optimization}, a comprehensive study is conducted to evaluate how risk in power systems can be modeled with (joint) chance-constrained programs. 

%Furthermore, our work does assume any specific technology for offering flexible capacity. Instead, we provide a general problem formulation in form of a \ac{DRJCCP} for aggregators or flexible providers to offer capacity of their stochastic flexible resources while adhering to specific Nordic market regulations.

%In this work, we also show how the \ac{TSO} can procure flexibility using a bi-level optimization problem with the inner problem being that of flexible demands offering capacity. We believe this is the first work to state such a problem. 

%In reference \cite{sheikhahmadi2021bi}, the authors use a bi-level problem to coordinate flexibility procurement between the \ac{DSO} and \ac{TSO}, but they do not look at how stochastic flexible resources react to changing market regulations. There is several studies on \ac{TSO}-\ac{DSO} coordination \cite{givisiez2020review, jiang2022flexibility}, but no studies on \ac{TSO}-aggregator dynamics.

%\subsection{Paper organization}

The rest of the paper is organized as follows. Section II defines the $\rm{P90}$ requirement and provides the problem formulation. Section III presents numerical results. Finally, Section IV concludes the paper. 

%First, we describe our problem formulation of bidding into Nordic ancillary service markets for stochastic flexible loads. We also formally define the P90 requirement from the regulations and show how it naturally corresponds to a \ac{JCC} from which flexible providers can use to model and subsequently bid their flexibility. We also introduce the \ac{TSO} perspective of maximizing flexible capacity as a bi-level optimization problem. Second, we show results for a simulated case study of a portfolio of \acp{EV} that have the ability to adjust their power consumption. We describe the simulation setup and investigate how an aggregator can bid \ac{EV} portfolio flexibility for different conservativeness levels, i.e., Wasserstein distances in a \ac{DRJCCP}. We also show how the \ac{TSO}' procurement of flexibility changes for different violation frequencies and Wasserstein distances. Finally, we conclude the paper and discuss future work.

\vspace{2mm}
\section{Problem Formulation}\label{sec:problem-formulation}
\vspace{-1mm}
We first define the $\rm{P90}$ requirement of the Danish TSO, Energinet, and describe how stochastic flexible resources can participate in Nordic ancillary service markets with respect to such a requirement. We show how it corresponds to an optimization problem with a single joint chance constraint. We then proceed to make it distributionally robust with respect to the stochastic consumption/production baseline of  flexible assets, leading to a  \ac{DRJCCP}, which can be  reformulated in an exact manner (i.e., no approximation) to a \ac{MILP}. Lastly, we embed this \ac{DRJCCP} within a bi-level optimization where the outer problem represents the \ac{TSO} perspective, and the inner problem represents the aggregator perspective. We will solve this bi-level program with a heuristic since the inner problem includes binary variables and thereby is not convex. 

\vspace{2mm}
\subsection{Definition of the $\rm{P90}$ requirement}
\vspace{-1mm}
Energinet has introduced the $\rm{P90}$ requirement as follows (the text is borrowed from \cite{energinet}): 
%
%Energinet has released the innovative regulation of the P90 requirement that incentivizes participation of stochastic flexible loads in their ancillary service markets \cite{energinet}. It specifically introduces an allowed violation frequency of 10\% as stated below:
%
\begin{definition}[The $\rm{P90}$ requirement]\label{def:P90}
    \textit{``[...] This means, that the participant's prognosis, which must be approved by Energinet, evaluates that the probability is 10\% that the sold capacity is not available. This entails that there is a 90\% chance that the sold capacity or more is available. This is when the prognosis is assumed to be correct.
    The probability is then also 10\%, that the entire sold capacity is not available. If this were to happen, it does not entail that the sold capacity is not available at all, however just that a part of the total capacity is not available. The available part will with high probability be close to the sold capacity."}
\end{definition}

Generally, the reservation market for ancillary services in Denmark is cleared the day before delivery. The $\rm{P90}$ requirement in Definition \ref{def:P90} states that the bid (for a given hour) submitted in the day-ahead stage should be feasible in real time at least 90\% of the time. Thus, the requirement allows for aggregators to (partially) fail in their forecast of their available flexibility. The second part of Definition \ref{def:P90} states that the magnitude of violations is not allowed to be extreme, thus discouraging severe overbidding (reserve shortfall). Lastly, we note that there are additional requirements for flexible \ac{LER} units \cite{energinet} which we do not consider them here, but they can readily be included in all subsequent formulations.

We now show how the interpretation of the $\rm{P90}$ requirement naturally leads to an optimization problem for an aggregator maximizing its profit defined as a joint chance-constrained program.

\vspace{1mm}
\subsection{Bidding problem of an aggregator}
\vspace{-1mm}
Hereafter, without loss of generality, we consider a flexibility aggregator in the demand side, where upward flexibility can be obtained by a decrease in consumption. To embed Definition \ref{def:P90} in an optimization problem, we consider the aggregator maximizes its profit by offering flexible capacity into a Nordic ancillary service market with an hourly resolution:
\begin{subequations}\label{P90:General}
    \begin{align}
    & \max_{p_{h}^{\text{cap}}} \quad  \sum_h \lambda_h \ p_{h}^{\text{cap}} \label{P90:General:obj}                                                                                                                                               \\
    &\text{s.t.} \nonumber  \\
    & \hat{\mathbb{P}}  \Big( p_{h}^{\text{cap}} \leq P_{m}^{\text{B}}, \ \forall{m} \in \mathcal{M}_{h},  \forall{h} \in \mathcal{H}  \Big) \geq 1 - \epsilon, \label{P90:General:jcc}
\end{align}    
\end{subequations}
where $h$ and $m$ are indexes for hours and minutes, respectively. Therefore, $\mathcal{H}\!=\!\{1, 2,  \ldots, 24\}$ represents hours for a given day, whereas $\mathcal{M}\!=\!\{1, 2,  \ldots, 1440\}$ includes all minutes in the same day. Further, $ \mathcal{M}_{h}\!=\!\{h \times 60 + m \mid m \in \{0, 1, 2, \ldots, 59\}\}$ represents all minutes in hour $h$. The objective function 
\eqref{P90:General:obj} maximizes the profit given deterministic hourly price forecasts $\lambda_h$ (in DKK/kW). The variable $p_{h}^{\text{cap}}$ is the capacity reservation bid (in kW) to the ancillary service market for hour $h$. This is consistent with Nordic ancillary service markets where the TSOs procure reservation capacity with an hourly resolution. Note that the capacity bid could either represent symmetrical (i.e., in both up and down directions) or unidirectional flexibility, depending on the market in question. %\textcolor{red}{We also note that \ref{Definition} can apply on a \textit{second} resolution as depending on the type of ancillary service. In that case, \eqref{P90:General} can be readily be modified to include all seconds within each hour.}

The sole uncertainty lies in the forecast for the stochastic baseline power consumption, denoted as uncertain parameter $P_{m}^{\text{B}}$ for minute $m$, governed by the empirical distribution $\hat{\mathbb{P}}$. Recall that at the time of bidding, the aggregator is uncertain of its future baseline power and therefore its ability to deliver flexibility. For example, for a portfolio of \acp{EV}, their future consumption is uncertain, but most likely predictable to some degree. Note that \eqref{P90:General:jcc} is a joint chance constraint, which can be interpreted as follows: there is a probabilistic constraint for every minute $m$ within the bid for hour $h$, enforcing $p_{h}^{\text{cap}} \leq P_{m}^{\text{B}}$. There are 1440 constraints in total in such a form. Constraint \eqref{P90:General:jcc} enforces a probability of at least $1-\epsilon$ for jointly fulfilling all those 1440 constraints. Following the $\rm{P90}$ requirement, we set $\epsilon\!=\!0.10$, which is consistent with Definition \ref{def:P90}. Note that we implement the $\rm{P90}$ requirement for the entire day, meaning we let the aggregator overbid with a probability of up to 10\% over a day. Note also that we consider a minute-level resolution for the stochastic baseline, although a finer resolution will be more desirable but it could be more challenging to obtain such a forecast. It could also increase the computational burden.

%Constraint \eqref{P90:General:jcc} is classified as a \ac{JCC} as it must be satisfied concurrently for all minutes throughout a day.

%The variability in available flexibility for the subsequent day, denoted as $P_{m}^{\text{B}}(\xi)$, is governed by a probability distribution $\mathcal{P}$. 
The general way to solve (joint) chance-constrained programs like \eqref{P90:General} is to generate \textit{samples} of $P_{m}^{\text{B}}$ by utilizing either historical data or a predictive distribution. Regardless of the method chosen, the resultant distribution may not accurately represent the true underlying uncertainty of the future power baseline. Consequently, the objective in this work is \textit{not} to generate the most representative sample set or predictions of $P_{m}^{\text{B}}$, but rather to identify optimal decisions based on a \textit{given} sample set, addressing any potential mis-specification of the sample set.
Therefore, instead of assuming a single empirical distribution in \eqref{P90:General}, we define an ambiguity set of distributions within some distance to the empirical distribution. To do this, inspired by \cite{chen2022data}, we exploit the well-known type-1 Wasserstein distance between two distributions, defined as
\begin{align}\label{was}
    d_{\mathrm{W}}\left(\mathbb{P}_1, \mathbb{P}_2\right)=\inf _{\mathbb{P} \in \mathcal{P}\left(\mathbb{P}_1, \mathbb{P}_2\right)} \mathbb{E}_{\mathbb{P}}\left[\left\|\tilde{\boldsymbol{\xi}}_1-\tilde{\boldsymbol{\xi}}_2\right\|\right],
\end{align}
where $\tilde{\boldsymbol{\xi}}_1 \sim \mathbb{P}_1$, $\tilde{\boldsymbol{\xi}}_2 \sim \mathbb{P}_2$, $\|\cdot\|$ denotes a norm, $\mathcal{P}\left(\mathbb{P}_1, \mathbb{P}_2\right)$ represents  the set of all distributions with marginals $\mathbb{P}_1$ and $\mathbb{P}_2$, and 
 $d_{\mathrm{W}}\left(\mathbb{P}_1, \mathbb{P}_2\right)$ gives the distance between two distributions $\mathbb{P}_1$ and $\mathbb{P}_2$.
We can then build our Wasserstein ambiguity set   $\mathcal{F}(\theta)$ as a ball of given radius $\theta \geq 0$, centered at the empirical distribution $\hat{\mathbb{P}}$, including any distribution $\mathbb{P}$ whose Wasserstein distance to $\hat{\mathbb{P}}$ is not greater than $\theta$:
\begin{align}\label{amb-set}
    \mathcal{F}(\theta)=\left\{\mathbb{P} \in \mathcal{P} \mid d_{\mathrm{W}}(\mathbb{P}, \hat{\mathbb{P}}) \leq \theta\right\}.
\end{align}

When $\theta \approx 0$, the ambiguity set in \eqref{amb-set} contains distributions very close to the empirical distribution $\hat{\mathbb{P}}$. When $\theta\!>>\!0$, the ambiguity set contains distributions very dissimilar to the empirical distribution $\hat{\mathbb{P}}$, thus being more conservative. For example, a large $\theta$ considers many different baseline consumption profiles of a portfolio of \acp{EV}.

We now move from the joined chance-constrained program \eqref{P90:General} to the following \ac{DRJCCP}: 
\begin{subequations}\label{P90:General:DRJCC}
    \begin{align}
       & \max_{p_{h}^{\text{cap}}} \quad  \sum_h \lambda_h \ p_{h}^{\text{cap}}                                                                                                                                                                                                     \\
            &\text{s.t.} \nonumber  \\
    &           \mathbb{P}  \left( p_{h}^{\text{cap}} \leq P_{m}^{\text{B}}, \ \forall{m} \in \mathcal{M}_{h},  \forall{h} \in \mathcal{H}  \right) 
         \geq 1 - \epsilon \notag\\
        & \hspace{6.1cm} \forall{\mathbb{P}} \in \mathcal{F}(\theta). \label{P90:General:drjcc_c}
    \end{align}
\end{subequations}

Problem \eqref{P90:General:DRJCC} thus optimizes the bidding decision for an aggregator of flexible resources participating in a Nordic ancillary service under Definition \ref{def:P90} using a distributionally robust representation of its baseline power. Note that \eqref{P90:General:drjcc_c} can be equivalently re-written as 
    \begin{align}
    &     \min_{\mathbb{P} \in \mathcal{F}(\theta)} \quad       \mathbb{P}  \left( p_{h}^{\text{cap}} \leq P_{m}^{\text{B}}, \ \forall{m} \in \mathcal{M}_{h},  \forall{h} \in \mathcal{H}  \right) 
         \geq 1 - \epsilon, \label{P90:General:drjcc_c2}
    \end{align}
meaning it picks the worst-case distribution within the ambiguity set $\mathcal{F}(\theta)$ and ensures the $\rm{P90}$ requirement taking into account that distribution.

%It can be used by any aggregator or flexible provider bidding into Nordic ancillary service markets with stochastic flexibility, e.g., wind mills, \acp{PV}, electrolyzers, \acp{TCL}, \acp{EV}, etc.
\vspace{2mm}
\subsection{Tractable reformulation of \eqref{P90:General:DRJCC}} 
\vspace{-1mm}
Problem \eqref{P90:General:DRJCC} can be reformulated in an exact manner (no approximation) as follows \cite[Proposition 2]{chen2022data}:
%\eqref{P90:General} is generally intractable although analytical reformulations exist \cite{nemirovski2007convex}. However, Problem \eqref{P90:General:DRJCC} emits a tractable reformulation as given by \cite[Proposition 2]{chen2022data}:
%
\begin{subequations}\label{P90:General:DRJCC-tract}
    \begin{align}
      &  \max_{p_{h}^{\text{cap}}, q_i, s_i \geq 0, t} \quad  \sum_h \lambda_h \ p_{h}^{\text{cap}}                                                                                                                                    \\
                &\text{s.t.} \nonumber  \\
    & \epsilon |\mathcal{I}| t - \sum_{i } s_i \geq \theta |\mathcal{I}|                                                                                      \\
                                                            & P_{m,i}^{\text{B}} - p_{h}^{\text{cap}} + M q_i \geq t - s_i, \ \notag\\
                                                            & \hspace{3.5cm} \forall{m} \in \mathcal{M}_{h},  \forall{h} \in \mathcal{H},  \forall{i} \in \mathcal{I} \\
                                                            & M (1-q_i) \geq t - s_i,   \quad \forall{i}  \in \mathcal{I}                                                                                                                                            \\
                                                            & q_i \in \{0,1 \}, \quad \forall{i} \in \mathcal{I},
    \end{align}
\end{subequations}
where $i\!\in\!\mathcal{I}$ is the set of (training) samples, $|\mathcal{I}|$ is the number of samples, and $P_{m,i}^{\text{B}}$ is the sample $i$ for the realization of uncertain parameter $P_{m}^{\text{B}}$ in minute $m$. Problem \eqref{P90:General:DRJCC-tract} in a \ac{MILP} including a set of auxiliary variables, namely binary variables $q_i$, non-negative continuous variables $s_i$, and the free continuous variable $t$. 
%Problem \eqref{P90:General:DRJCC-tract} is a \ac{MILP} with binary variables, $q_{i}$, that indicate a violation for sample $i \in \mathcal{I}$. As mentioned, $\theta$ specifies how far samples from the empirical distribution can be transported, as shown in \eqref{amb-set}. The 
Parameter $M$ is called the Big-M value, which is a large enough positive constant. This reformulation is quite sensitive to the magnitude of $M$ which becomes evident when $\theta\!=\!0$. Here, $M$ constrains the maximum violation of the bid, causing the empirical distribution to lose its significance \cite[Theorem 2, Remark 1]{chen2022data}.

It is not immediately obvious what value the parameter $\theta$ should take. Recall, the aggregator should set $\theta$ \textit{before} solving \eqref{P90:General:DRJCC-tract}. A high value for $\theta$ leads to more conservative capacity bids, a lower profit, but more security of delivery. On the contrary, a small value for $\theta$ resembles the empirical distribution and finds the highest possible capacity bids and profit, but this might prove sensitive to non-stationarity in the baseline power $P_{m}^{\text{B}}$. An option is to solve \eqref{P90:General:DRJCC-tract} for a predefined set of $\theta$ values and choose the smallest, feasible value of $\theta$, as suggested by \cite[Section 3.2]{chen2022data}.

Alternatively, the \ac{TSO} could also prescribe a value for $\theta$ according to its needs as discussed in the next subsection.

\vspace{2mm}
\subsection{Tuning problem of a \ac{TSO}}
\vspace{-1mm}
We discuss how a TSO can tune not only $\epsilon$ but also $\theta$. 
The \ac{TSO} is primarily interested in procuring sufficient amount of reserve capacity to reliably balance the power grid. This is becoming an increasingly difficult decision-making problem for the \ac{TSO} with more intermittent renewable generation in the power grid, but also with a pressure to not procure (too much) fossil-fuel based reserve capacity. For these reasons, Definition \ref{def:P90} was invented by Energinet in the first place to allow stochastic flexible resources to bid into ancillary service markets. However, the violation frequency of $\epsilon\!=\!0.10$ in Definition \ref{def:P90} is rather arbitrary and one could imagine a higher value for $\epsilon$ may yield more flexible capacity but with less certainty of delivery. Therefore, we formulate an optimization problem for the \ac{TSO} to determine $\epsilon$ in response to how the set of aggregators, indexed by $d\!\in\!\mathcal{D}$, behaves according to \eqref{P90:General:DRJCC-tract}. Furthermore, we assume the \ac{TSO} can prescribe a degree of conservativeness upon the aggregators by pre-setting $\theta$. This ends up in a bi-level program enabling the TSO to jointly determine values for $\epsilon$ and $\theta$:
%, and that it ignores the cost of procurement. 
%
%Then \ac{TSO} capacity procurement is described by the following bi-level optimization problem:
%
\begin{subequations}\label{P90:TSO}
    \begin{align}
        & \max_{\epsilon, \theta, \nu_{m,h,i,d}} \quad \sum_{h,d} p_{h,d}^{\text{cap}} -  \frac{1}{|\mathcal{I}|}\sum_{m,h,i,d} \nu_{m,h,i,d}     \label{P90:TSO:outer1}                                                                                                                                                                                                  \\
        & \text{s.t.} \nonumber \\
        & \nu_{m,h,i,d} = \left(p_{h,d}^{\text{cap}} - P_{m,d,i}^{\text{B}} \right)^{+}, \notag\\ &\hspace{2.5cm}\forall{m} \in \mathcal{M}_{h},  \forall{h} \in \mathcal{H}, \forall{i} \in \mathcal{I}, \forall{d} \in \mathcal{D} \label{P90:TSO:outer2}   \\
        & \text{Problem} \thinspace \eqref{P90:General:DRJCC-tract}, \quad \forall{d} \in \mathcal{D}, \label{P90:TSO:inner} 
    \end{align}
\end{subequations}
where $(.)^{+}\!=\!\text{max} \{.,0\}.$
In the inner optimization problem \eqref{P90:TSO:inner}, each aggregator $d$ maximizes its profits as per \eqref{P90:General:DRJCC-tract} and bids hourly reserve capacities $p_{h,d}^{\text{cap}}$. In the outer optimization problem \eqref{P90:TSO:outer1}-\eqref{P90:TSO:outer2}, the \ac{TSO} optimizes over $\epsilon$ and $\theta$ which are treated as parameters for aggregators in the inner optimization. The objective of the \ac{TSO} is to incentivize aggregators to bid reserve capacities as \textit{reliable} as possible, given that aggregators maximize their profit by imposing distributional robustness on their uncertain baseline power. Variables $\nu$ represent unavailable reserve capacity from aggregators and is subtracted (on average over $m$ and $i$) from the total reserve capacity over a day. Note that  \eqref{P90:TSO:outer1} considers reliability only and discards any financial incentive for the TSO. One may re-write this objective function in a way that it minimizes the reserve procurement cost of the TSO accounting for the cost associated to the lack of reliability. 

Recall the inner problem \eqref{P90:TSO:inner} for each aggregator $d$ is a MILP and therefore non-convex. This means one cannot solve the bi-level problem \eqref{P90:TSO}, as common in the literature \cite{pozo}, by replacing \eqref{P90:TSO:inner} with its optimality conditions. We therefore solve it in a heuristic way by pre-defining a finite set of potential values for $\epsilon$ and $\theta$, solve inner problem for each combination of $(\epsilon$, $\theta$), and find their optimal values within the available  options from the TSO's perspective. 

\vspace{2mm}
\section{Numerical Results}
\vspace{-1mm}
Through our numerical simulations, we first explore 
how an aggregator of flexible \acp{EV} can offer reserve capacity by using \eqref{P90:General:DRJCC-tract}. For this, we conduct a sensitivity analysis with respect to $\theta$ to explore impacts of the conservativeness level. We then take the \ac{TSO} perspective and solve \eqref{P90:TSO} to  maximize the total reserve capacity offered while minimizing overbidding (reserve shortfall). All source codes are publicly available in \cite{code}.

%, with respect to \ac{IS} and \ac{OOS} performance, and \textit{(ii)} how a \ac{TSO} can use \eqref{P90:TSO} to tune the violation frequency $\epsilon$ and the conservativeness level $\theta$ to maximize the total reserve capacity offered while minimizing overbidding (reserve shortfall).   

%First, we briefly introduce the simulation setup, and explain how \ac{EV} consumption is modelled. We then present the main findings of \textit{(i)} and \textit{(ii)}.

\begin{figure}[t]
    \centering
    \includegraphics[width=\columnwidth]{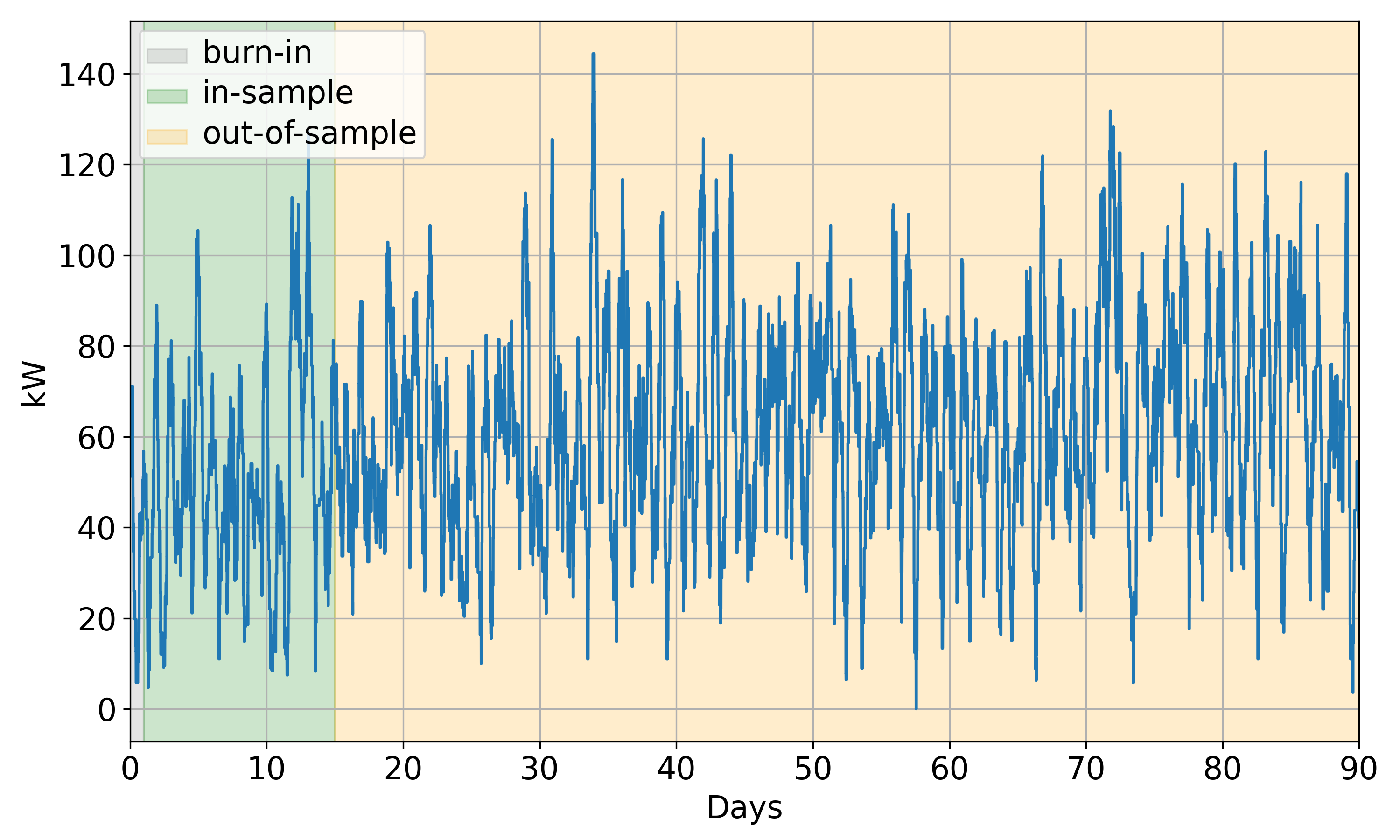}
    \caption{\small{Input data: Simulation of \ac{EV} power consumption for 90 days. First day is a burn-in period, green is the in-sample period (constructing the empirical distribution), and yellow is the out-of-sample period.}}
    \label{fig:drjcc_raw}
\end{figure}

\vspace{2mm}
\subsection{Input data}\label{sec:sim-setup}
\vspace{-1mm}
%As mentioned, Definition \ref{def:P90} and Problem \eqref{P90:General:DRJCC-tract} is agnostic towards any stochastic flexible resource. For this case study, we consider an aggregator of \acp{EV} with flexibility to turn their power consumption up or down, and $p_{m}^{\text{B}}(\xi)$ represents the uncertain \ac{EV} portfolio baseline power consumption.

Fig. \ref{fig:drjcc_raw} illustrates a simulation for the consumption of a portfolio of 20 \acp{EV} over 90 days. Weekend days are simulated to have higher consumptions on average, and the same is the case for evenings. To introduce a degree of  non-stationarity, the average charge time grows proportional to the square root of time, consequently making a positive drift in the average portfolio consumption. This scenario could symbolize, for instance, the effect of colder weather, where \acp{EV} require longer charging periods due to increased energy demand.

We call the first 15 days of Fig. \ref{fig:drjcc_raw}, highlighted in green, as in-sample period. We use the consumption data in this period to construct the empirical distribution which is an input data for optimization models \eqref{P90:General:DRJCC-tract} and \eqref{P90:TSO}. We then use the remaining 75 days, called the out-of-sample period highlighted in yellow, for evaluating the quality of bids obtained by models \eqref{P90:General:DRJCC-tract} and \eqref{P90:TSO}.

%in Fig. \ref{fig:drjcc_raw} defines the empirical distribution to be used when solving \eqref{P90:General:DRJCC-tract} and \eqref{P90:TSO}. The resulting bids are then evaluated against the out-of-sample (OOS) period data. 

This is of course a stylized way of intentionally highlighting the impact of distributional robustness, but it generalizes to any situation where any flexible provider might have mis-specification in its empirical (or predictive) distribution for any technology with stochastic baseline power.

\vspace{2mm}
\subsection{Optimal reserve capacity bids of the aggregator}
\begin{figure}[!t]
    \centering
    \begin{adjustbox}{width=\columnwidth}
        \input{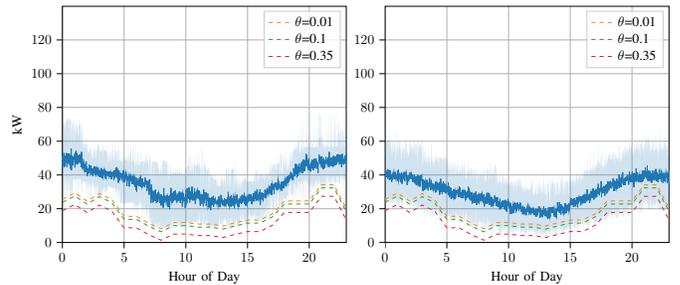}
    \end{adjustbox}
    \caption{\small{ As input data, the blue curve shows the stochastic consumption of the portfolio on 20 EVs over a day, representing the mean (dark blue) and 10-90\% quantiles (light blue). The left plot shows the \textit{in-sample} historical consumption data which are already represented in the green area of Fig. \ref{fig:drjcc_raw}. The right plot includes the \textit{out-of-sample} consumption data coming from the yellow part of that figure. As outputs of model \eqref{P90:General:DRJCC-tract} given in-sample data, the three dashed line curves (identical in both plots) show the optimal  reserve capacity bids for three values of $\theta = \{0.01, 0.1, 0.35\}$ representing different levels of conservativeness.}}
    \label{fig:drjcc_bids}
\end{figure}

\vspace{-1mm}
We aim to derive optimal reserve capacity bids for every hour of the next day using historical consumption data represented in Fig. \ref{fig:drjcc_raw}. We use this historical data to represent the hourly uncertain consumption of the next day, illustrated by the blue curve in Fig. \ref{fig:drjcc_bids}, where the left (right) plot corresponds to the in-sample (out-of-sample) period. The dark blue curve gives the mean, whereas the light blue shows the 10-90\% quantiles. 

For optimal bidding in every hour, we take $|\mathcal{I}| = 30$ samples from  in-sample data. The three dashed line curves in Fig. \ref{fig:drjcc_bids} (identical curves in both plots) show the optimal upward reserve capacity bids for  three values of $\theta = \{0.01, 0.1, 0.35\}$ representing different levels of conservativeness. 
%shows bids from an aggregator of 20 \acp{EV} and with $|\mathcal{I}| = 30$ with respect to their \ac{IS} available flexibility\footnote{We note that in this study, the uncertain future baseline power consumption corresponds to the available flexibility.} (as shown in blue) for different values of conservativeness $\theta$.
Starting with the in-sample data (left plot), all three curves of hourly bids adhere to Definition \ref{def:P90}. For $\theta = 0.01$, the bids almost represent the 90\% quantile, i.e., the maximum allowed frequency of reserve shortfall. By increasing the value of $\theta$, the aggregator becomes more conservative and therefore offers comparatively lower reserve quantities. However, as observed from the out-of-sample results (right plot), the bids for $\theta = 0.35$ are the only ones that (barely) adhere to the $\rm{P90}$ requirement. This is obviously due to the non-stationarity introduced in Fig. \ref{fig:drjcc_raw}. This highlights how important it is for the EV aggregator to behave conservatively and be ambiguity-averse when bidding in an non-stationary environment. One can argue it could be the case  
%Clearly, a low $\theta$ is risky in case the empirical distribution of available flexibility is just a little mis-specified. This might very well be the case 
for other technologies as well, e.g., for wind and solar power units or aggregators of other types of stochastic flexible demands who are prone to sudden disturbances. 
%This highlights the awareness aggregators should have when bidding, but also how Definition \ref{def:P90} can be exploited to artificially allow for too high bids. 

%This essentially corresponds to \textit{overfitting} to the empirical distribution, a term common in the machine learning literature \cite{bishop2006pattern}.

\begin{figure}[!t]
    \centering
    \includegraphics[width=0.95\columnwidth]{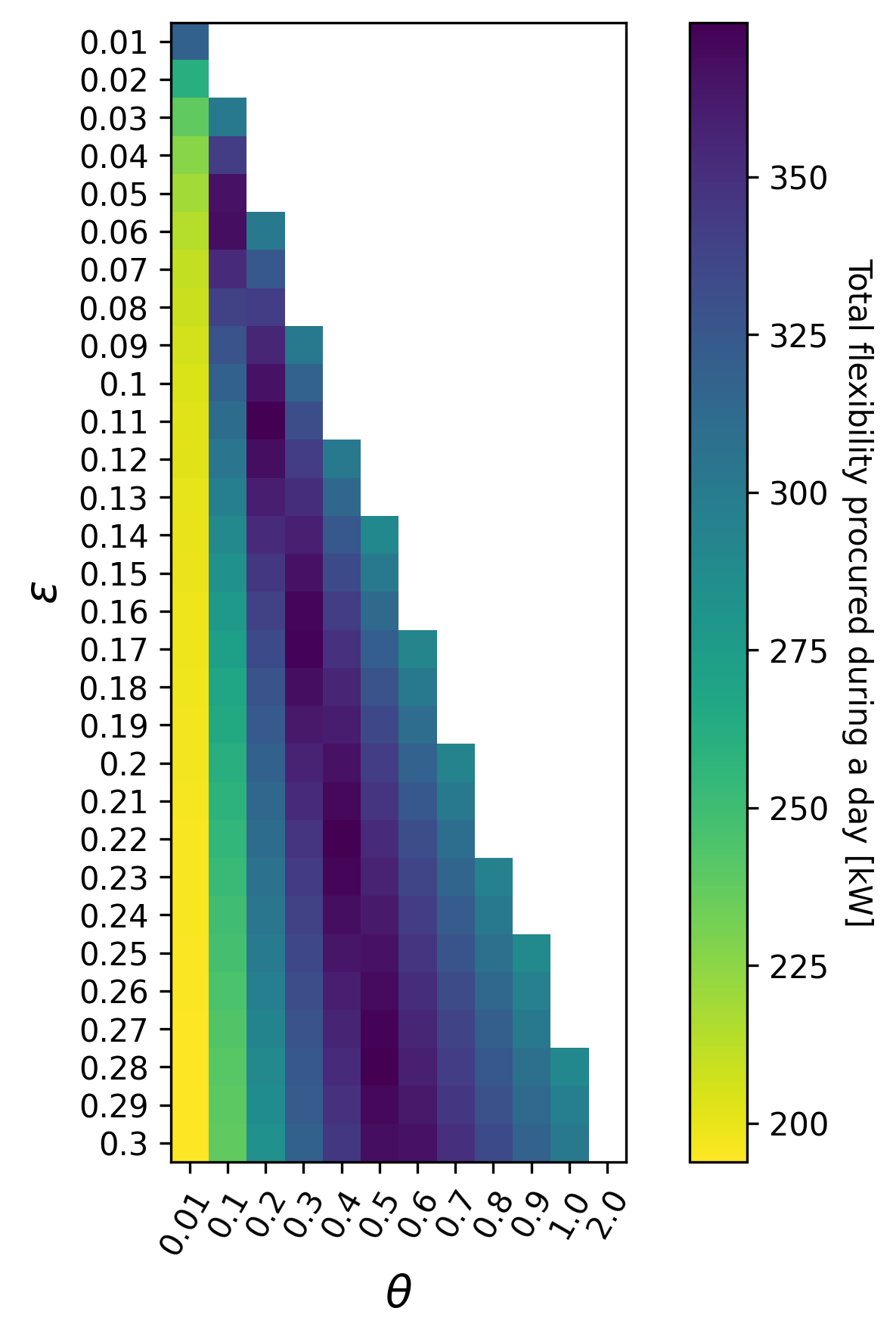}
    \caption{\small{Total reserve capacity procurement of the \ac{TSO} over a day from the aggregator of 20 EVs for different values of $\epsilon$ and $\theta$. Each $(\epsilon, \theta)$ has been solved using \eqref{P90:TSO} using a grid search on in-sample data. The optimal value obtained by this grid search heuristic is $(\epsilon, \theta) = (0.11, 0.2)$. \vspace{-2mm}}}
    \label{fig:tso}
\end{figure}

\vspace{2mm}
\subsection{Reserve capacity procurement by the \ac{TSO}}\label{sec:motivation}
\vspace{-1mm}
Fig. \ref{fig:tso} shows the total amount of reserve capacity procurement of the \ac{TSO} over a day from the aggregator of 20 EVs for different values of $\epsilon$ and $\theta$. The main observation is that a high $\epsilon$, i.e., allowing a high frequency of reserve shortfall, yields more available flexibility for the \ac{TSO} \textit{only if} $\theta$ is also high. This is due to the variables $\nu$ in \eqref{P90:TSO} which penalize the reserve shortfall. Interestingly, we also observe a path of optimal $(\epsilon, \theta)$ starting from $(0.01, 0.01)$ to $(0.30, 0.05)$. This demonstrates that reserve shortfall for high $\epsilon$ can be compensated by more conservative decision making while maintaining a high reserve capacity procurement for the \ac{TSO}.

\vspace{2mm}
\section{Conclusion}
\vspace{-1mm}
Given the $\rm{P90}$ requirement of Energinet, this paper illustrated how stochastic flexible resources can participate in Nordic ancillary service markets using a \ac{DRJCCP} for offering reserve capacity. We showed how a tractable formulation of the model allows for more robust bidding in non-stationary environments of the uncertain baseline power consumption. Furthermore, the perspective of the \ac{TSO} was investigated with respect to its reserve capacity procurement from such stochastic flexible resources. It was shown how an increased allowance for the probability of reserve shortfall should be offset by a corresponding increase in conservativeness of bidding using the measure of the Wasserstein distance in the \ac{DRJCCP}. These findings were exemplified using a simulated case study of an aggregator of \acp{EV} bidding their flexibility into an ancillary service with negligible energy delivery.

For future work, it is of great interest to investigate the impact of a heterogeneous portfolio of stochastic flexible resources with respect to the \ac{TSO} procurement. As such, one could expect that flexible resources of different technology also respond differently when the \ac{TSO} prescribes allowed probability of shortfall and conservativeness. Moreover, the \ac{TSO} procurement model in this work ignores prices for ancillary services and penalty prices for reserve shortfall. Both are important from a system and societal perspective and should be included in future work. Lastly, an increase in the volume of flexible resources bidding in ancillary service markets lowers prices which will be offset by more uncertain supply. This trade-off could be interesting from the \ac{TSO} perspective, and should be  further studied.

\vspace{2mm}

% \section*{Acknowledgement}


\bibliographystyle{IEEEtran}

\bibliography{tex/bibliography/Bibliography}
%\bibliography{tex/bibliography/Bibliography}

\vfill

\end{document}